\documentclass{PoS}

\title{Extraction of Neutrino Flux from the Inclusive Muon Cross Section}

\ShortTitle{Extraction of Neutrino Flux from the Inclusive Muon Cross Section}

\author{\speaker{Tomoya Murata}\\%
        Osaka University\\
        E-mail: \email{murata@kern.phys.sci.osaka-u.ac.jp}}

\author{Toru Sato\\
        Osaka University\\
        E-mail: \email{tsato@phys.sci.osaka-u.ac.jp}}

\abstract{ 
We have studied a method to extract neutrino flux from the data 
of neutrino-nucleus reaction by using maximum entropy method.
We demonstrate a promising example to extract neutrino flux
from the inclusive cross section of muon production
without selecting a particular reaction process such as quasi-elastic
nucleon knockout.}

\FullConference{16th International Workshop on Neutrino Factories and Future Neutrino Beam Facilities\\
		 25 -30 August, 2014\\
		 University of Glasgow, United Kingdom}

\newcommand{\eqbegin}{\begin{eqnarray}}
\newcommand{\eqend}{\end{eqnarray}}
\begin{document}

\section{Introduction}

A precise description of the neutrino-nucleus interaction 
is crucial for the study the neutrino properties
such as CP violating phase and mass hierarchy\cite{hyper-k,lbne,lbno,bgadra}
when one extracts the neutrino flux from the neutrino data.
The  quasi-elastic (QE) reaction is the key mechanism
to reconstruct the neutrino flux from the data of neutrino-nucleus reaction.
Neutrino energy($E_\nu$) can be determined from the muon 
scattering angle($\theta$) and
energy($E_\mu$) by using the kinematics of the reaction of free nucleon at rest.
\begin{eqnarray}
E_\nu & = & \frac{E_\mu M_N - m_\mu^2/2}{M_N - E_\mu + p_\mu \cos\theta}
\label{eq-enu}
\end{eqnarray}
Since the reaction takes place inside nuclear many body system,
various initial and final state interactions can spoil the 
useful relation of Eq. (\ref{eq-enu}).
The observed quasi-elastic like events are
mixture of the various reaction mechanism, which
contribute the reaction amplitudes and, in principle,
interference among various mechanism exists.
The central issue of the current theoretical research of neutrino-nucleus reaction
is to improve the accuracy of the estimation of
various nuclear effects\cite{alvarez14,garvey14} by testing the model from
the neutrino data.

Because of the rather wide energy band of the neutrino flux
in the current long baseline experiments,
the observed cross section of neutrino reaction
($<\sigma>$) is a result of neutrino flux($\Phi_\nu$) average 
of the cross section ($\sigma$).
\begin{eqnarray}
<\sigma> & = & \int dE_\nu \sigma(E_\nu) \Phi_\nu(E_\nu)
\end{eqnarray}
One can regard the extraction of neutrino flux $\Phi_\nu$ from the data $<\sigma>$
as an inverse problem to obtain $\Phi_\nu$ from $<\sigma>$
provided the cross section $\sigma$ is well
under control. In this report we will examine a method 
to solve the inverse problem, where the neutrino flux 
can be extracted without using the formula (\ref{eq-enu}) nor the separation
of 'true' QE events. For this purpose, we use the maximum entropy method (MEM)
which are widely used in the field of
condensed matter physics and Lattice QCD\cite{silver,gubern,asakawa2001}.

We briefly outline the application of MEM to extract
neutrino flux from data in section 2.
To examine the usefulness of the approach, 
we generate pseudo data of neutrino reaction.
The pseudo data are generated by using 
a model of neutrino reaction and neutrino flux,
which are explained in section 3.
Here, as an example of the pseudo data,
we use double differential cross section of muon
for the inclusive neutrino nucleus reaction.
Results on how well the neutrino flux can be extracted from the 
pseudo 'data' without assuming QE mechanism is shown in section 4.

\section{Maximum Entropy Method}

In this section we briefly explain the maximum entropy method. 
The most plausible neutrino flux ($\Phi_\nu^{MEM}$) from the experimental data
based on the Bayes' theorem
 is given by the functional integral of the neutrino flux $\Phi$
\begin{eqnarray}
 \Phi_\nu^{MEM} & = & \int [d\Phi] \Phi P[\Phi|\bar{D},I],
\end{eqnarray}
where $P[\Phi|\bar{D},I]$ is a conditional probability of neutrino flux $\Phi$ 
for given data of neutrino reaction $\bar{D} = \{<\sigma>_{exp}\}$ and the prior
information on the neutrino flux $I$. $I$ is called as default model of neutrino flux.
Introducing auxiliary variable $\alpha$, we can write the above formula as
\begin{eqnarray}
 \Phi_\nu^{MEM} & = & \int d\alpha  \Phi_\nu^\alpha P[\alpha|\bar{D},I], \label{eq-phi}
\end{eqnarray}
where $\Phi_\nu^\alpha$ is neutrino flux for given $\alpha$, data $\bar{D}$ and prior information $I$ given as,
\begin{eqnarray}
 \Phi_\nu^\alpha & = & \int [d\Phi]  \Phi P[\Phi|\alpha,\bar{D},I] 
            \propto  \int [d\Phi]  \Phi P[\bar{D}|\Phi,\alpha,I] P[\Phi|\alpha,I].
\end{eqnarray}
Here $P[\bar{D}|\Phi,\alpha,I]$ and  $P[\Phi|\alpha,I]$ are called as likefood 
function and prior
probability, respectively. The likefood function is written as
\begin{eqnarray}
P[\bar{D}|\Phi,I] & = & \frac{1}{Z_\chi}\exp\left(-\frac{1}{2}\chi^2\right),
\end{eqnarray}
with
\begin{eqnarray}
  \chi^2 & =& \sum_{l=1}\frac{(\bar{D_l}-D_l)^2}{\sigma_l^2} \nonumber \\
  Z_\chi & = & \prod_{l=1} \sqrt{2\pi \sigma_l^2}.
\end{eqnarray}
Here $\bar{D}_l$ and $D_l$ are cross sections $<\sigma>_{exp}$ and   $<\sigma>$.

The prior probability can be expressed as
\begin{eqnarray}
P[\Phi|\alpha,I] = \frac{1}{Z_S} \exp\left(\alpha S\right),
\end{eqnarray}
where S is the Shannon-Jaynes entropy,
\begin{eqnarray}
S = \sum_i \left(\Phi_i - m_i - \Phi_i \ln\frac{\Phi_i}{m_i} \right).
\end{eqnarray}
$m_i$ is default model of  neutrino flux at neutrino energy $E_i$.
Combining the likefood function and the prior probability,
neutrino flux $\Phi_\nu^\alpha$ is the maximum probability of
$\displaystyle P[\Phi|\alpha,\bar{D},I] \propto e^{Q(\Phi)}$
with $Q(\Phi) = \alpha S - \frac{1}{2}\chi^2$.

Finally the $\alpha$ dependent flux is integrated with
the probability $P[\alpha|\bar{D},I]$, which has a sharp peak 
as a function of $\alpha$ written as
\begin{eqnarray}
P[\alpha|\bar{D},I] & = & \int [d\Phi] P[\Phi \alpha|\bar{D},I]
  \propto P[\alpha|I] \int [d\Phi] \frac{1}{Z_S Z_\chi}e^{Q(\Phi)}
\end{eqnarray}
Assuming constant $P[\alpha|I]$, 
we obtain the neutrino flux $\Phi_\nu^{MEM}$ by integrating $\alpha$ around 
the shark peak of $P[\alpha|\bar{D},I]$ in Eq. \ref{eq-phi}.

One can also estimate the errors of the extracted neutrino flux $\Phi_\nu^{MEM}$
within MEM.
For given $\alpha$, the average flux $<\Phi_\alpha>_S$ in energy region $S$
of neutrino  is defined as
\begin{eqnarray}
\left<  \Phi_\nu^\alpha  \right>_S 
&=&  \frac{\int_S dE_\nu \Phi_\nu^\alpha(E_\nu)}
{\int_S dE_\nu}.
\end{eqnarray}
The covariance of $\left<  \Phi_\nu^\alpha  \right>_S$ can be written as
\begin{eqnarray}
\left< (\delta \Phi_\nu^\alpha)^2 \right>_S
&=& \int_{S^2}dE_\nu dE'_\nu \delta \Phi(E_\nu) \Phi(E'_\nu) P[\Phi|\bar{D},\alpha,I]/\int_{S^2}dE_\nu dE'_\nu \nonumber \\
&\simeq& -\int_{S^2} dE_\nu dE'_\nu \left( \frac{\delta^2 Q}{\delta\Phi(E_\nu)\delta\Phi(E'_\nu)}
  \right)^{-1}_{\Phi=\Phi_\alpha}/ \int_{S^2}dE_\nu dE'_\nu.
\end{eqnarray}
Then the error of the extracted flux is obtained by integrating over $\alpha$ as 
\eqbegin
\left< (\delta\Phi_\nu^{MEM})^2  \right>_S 
= \int d\alpha \left< (\delta\Phi_\nu^{\alpha})^2 \right>_S P[\alpha | \bar{D},I].
\eqend

\section{Model of inclusive neutrino-nucleus cross section and neutrino flux}

To examine the method outlined in the previous section, 
we construct pseudo data of neutrino-nucleus reaction by using a model
of neutrino-nucleus reaction and neutrino flux.
The neutrino flux is obtained by applying the formula of two-flavor oscillation
as
\eqbegin
\Phi(E_\nu) &  = & \left(1-\sin^2 2\theta_{23} \sin^2 \left(\frac{\Delta m_{23} L}{E_\nu}\right)\right)\Phi_\nu^0(E_\nu) \label{eq-fluxosc}
\eqend 
Here we used $\Delta m_{23}^2 = 2.5\times 10^{-3} {\rm eV^2}$, $\sin^2 2\theta_{23} = 1.0$
and $L = 295 km$. The initial flux $\Phi_\nu^0(E_\nu)$ is taken from near detector 
neutrino flux of T2K experiment Ref. \cite{t2kflux}.
\begin{figure}[h]
    \begin{center}
      \includegraphics[width=8cm]{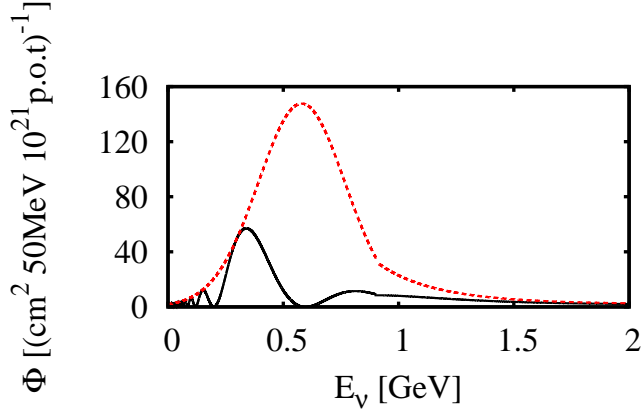}
    \end{center}
  \caption{Neutrino flux. The solid(black) and
the dashed(red) lines show the neutrino flux $\Phi_\nu$ and
 $\Phi_\nu^0$.}
  \label{fig:flux-nd}
\end{figure}
The solid(black) curve in Fig. 1 is the neutrino flux we use to calculate
 the pseudo data,
while the original flux $\Phi_\nu^0$ is shown in dashed(red) curve for comparison.

As a observable to test the method of flux extraction,
we use double differential cross section of muon
of neutrino induced inclusive reaction on $^{12}C$,
$\nu_\mu + ^{12}C \rightarrow \mu^- + X$
\eqbegin
 \frac{d^2 \sigma}
{dE_\mu d\Omega_\mu}.
\eqend
In the energy region of neutrino flux in Fig. 1,
the main reaction mechanism is quasi elastic scattering and 
pion production through the $\Delta(1232)$ excitation.
The quasi elastic nucleon knockout process is calculated using
modified Fermi-gass model which incorporates the spectral function
\cite{benhar94,benhar2005}. For the delta excitation region
we use the neutrino pion production amplitudes calculated in a formalism in which 
the resonance contributions and the background amplitudes are 
treated on the same footing\cite{sato2003}. 
Detail on the reaction model is described in Ref. \cite{szczer}.
The model of nuclear reaction is rather simple without 
taking into account the final state interaction,
however the reliability of our formalism is tested against the electron
nucleus scattering cross sections in the same theoretical framework; 
the calculated cross sections agree reasonably well with the existing data.
Therefore, we believe it is good enough for the purpose of
testing the method of neutrino flux extraction.

We now are able to  calculate the flux averaged double differential cross section 
$ d^2 \sigma/dE_\mu d\Omega_\mu$
from the model of neutrino-nucleus reaction and neutrino flux
$\Phi_\nu$ in Eq. (3.1).
\eqbegin
\left< \frac{d^2\sigma}
{dE_\mu d\Omega_\mu}\right>
=\int dE_\nu \frac{d^2 \sigma}
{dE_\mu d\Omega_\mu} \Phi(E_\nu),
\label{eq:crs-ave}
\eqend

The pseudo data of double differential cross section is obtained as following.
We assume muon energy resolution is 100MeV and evaluate the 
average of cross section as,
\begin{eqnarray}
\left< \overline{\frac{d^2\sigma}
{dE_\mu d\Omega_\mu}}\right> = 
\int_{E_{1}}^{E_2} dE_\mu \frac{1}{E_2 - E_1}
\left< \frac{d^2\sigma}
{dE_\mu d\Omega_\mu}\right>
\end{eqnarray}
We assume Gaussian distribution of the data
with the mean given by Eq. (3.4) and
the standard deviation by the 
 10\% of the value of flux averaged double differential cross section 
at $E_\mu$=1GeV.
Fig. 2 shows the obtained pseudo data of the muon energy distribution
 at the muon scattering angle $\theta_\mu=10^\circ$.
\begin{figure}[h]
    \begin{center}
      \includegraphics[width=8cm]{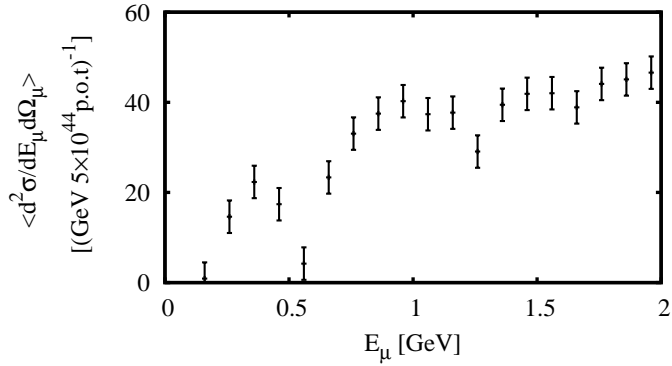}
    \end{center}
  \caption{Pseudo data of neutrino induced 
inclusive muon production cross section on $^{12}C$.
The scattering angle of the muon is  $\theta_\mu=10^\circ$.}
  \label{fig:pseudo-data}
\end{figure}

\section{Results and discussion}

We extract neutrino flux from the 19 points of pseudo data 
of the muon energy distribution shown in Fig. 2.
They give  $\bar{D}_l$ and $\sigma_l$ of Eq. (2.5).  $D_l$ is calculated
from the theoretical cross section of Eq. (3.2).

At first, we study the effects of  the default model $I$ on
the extracted neutrino flux.
We have examined two default models.
The one is constant flux  and the other
is initial flux  $\Phi_\nu^0$ shown in the dashed(red) 
and solid(black) curve in the left panel of Fig. 3.
\begin{figure}[h]
    \begin{center}
      \includegraphics[width=6cm]{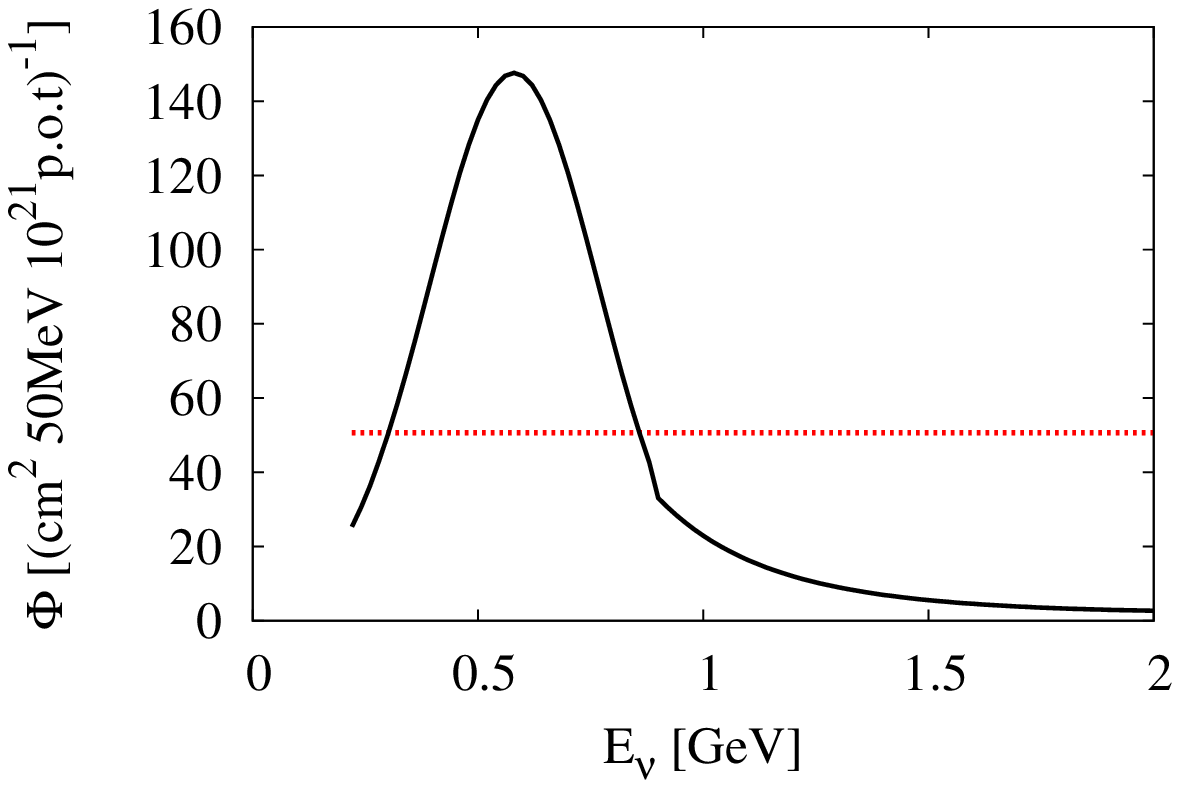}
      \includegraphics[width=6cm]{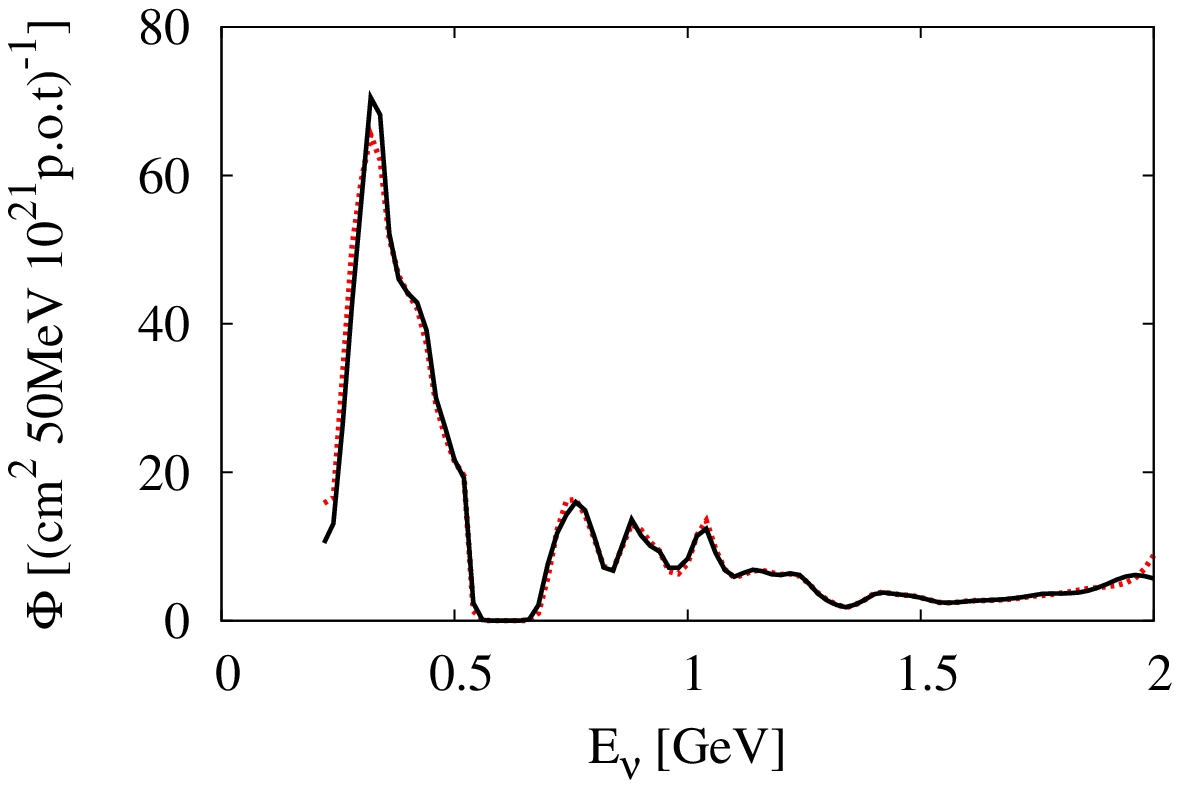}
    \end{center}
  \caption{Default model dependence of extracted neutrino flux.
The solid(black) and dashed(red) curves 
in the left panel show 
 $\Phi_\nu^0$ and constant flux
as default model. The extracted flux $\Phi_\nu^{MEM}$ 
corresponding to each default model is shown in the right panel.
}
  \label{fig:reconstructed-model-const}
\end{figure}
The extracted flux using each default model is shown in the
right panel of Fig. 3. One can see that, in this case, the extracted flux
is very stable against the choice of the default model.

We now compare the extracted flux $\Phi_\nu^{MEM}$(solid curve) 
with the expected 'exact' neutrino flux $\Phi_\nu$(dashed curve) in Fig. 4.
Extracted flux agrees well with the 'exact' flux over the wide muon energy region.
The extraction of the flux is only valid on the energy above the
reaction threshold and the flux extraction did not work well near the threshold
energy. We noticed that
the extracted flux has structures in the $0.7 < E_\nu < 1$ GeV region.
\begin{figure}[h]
    \begin{center}
      \includegraphics[width=8cm]{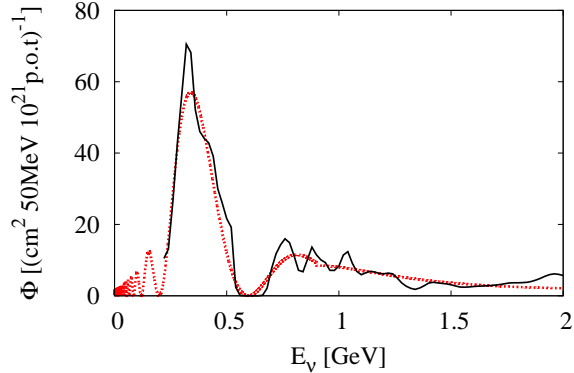}
    \end{center}
  \caption{Comparison of extracted flux $\Phi_\nu^{MEM}$(solid curve) and
 'true' flux $\Phi_\nu$(dashed curve).}
  \label{fig:reconstructed-flux10-4}
\end{figure}
One can judge the structure is real or not by estimating the
error of extracted flux.
In Table 1, the neutrino flux averaged
in the peak region ($0.28 < E_\nu < 0.42$ GeV),
dip region ($0.52 < E_\nu < 0.70$ GeV) and higher energy regions
($0.72 < E_\nu < 0.82$ GeV), ($0.86 < E_\nu < 0.94$ GeV) and ($1.00 <
E_\nu < 1.08$ GeV)
 are given.
The second raw shows the calculated 'exact' value of the average flux
and the third raw shows the flux obtained from  MEM together 
with the error calculated from Eq. (2.11).
The central values of the $\Phi_\nu^{MEM}$ agree well 
with the 'exact' ones within the errors in all the energy regions.
The size of the oscillation of $\Phi_\nu^{MEM}$ seen in 
$0.7 < E_\nu < 1$ GeV is smaller than the estimated errors in that region.
Therefore one can conclude that we sould not take the oscillation 
of the obtained flux seriously from the error estimation 
even without knowing the 'exact' flux.

\begin{table}[h]
\begin{center}
\begin{tabular}{cccccc}\hline
$E_\nu$(GeV)  & $0.28 - 0.42$   & $0.52-0.70$     & 0.72-0.82       & 0.86-0.94 & 1.00-1.08 \\
Exact        & $53.5$          &  $3.9$          & $10.5$          & $10.9$   &  $8.7$ \\
MEM          & $56.5 \pm  6.5$ &  $5.1 \pm  1.7$ & $14.0 \pm 3.7$  & $11.6 \pm 3.7$ & $10.7 \pm 1.9$ \\ \hline
\end{tabular}
\end{center}
\caption{Energy averaged neutrino flux.}
\end{table}

In summary, we have examined MEM to extract the neutrino flux from 
the inclusive muon production cross section.
The results show the proposed method can be useful alternative tool 
to extract the neutrino flux provided accurate theoretical cross section is known.
Further work to find the optimal observables for the efficient extraction
of flux with the refined model of neutrino-nucleus reaction should be done
in near future.

The authors would like thank to Drs. M. Kitazawa and S. Nakamura
for illuminating discussions.
This work was supported by JSPS KAKENHI Grant Nos. 24540273 and 25105010.

\end{document}